\documentclass[prd, amsmath, aps, floats,floatfix,amssymb, superscriptaddress, nofootinbib,preprintnumbers,12pt]{revtex4}

\usepackage[dvips]{graphicx}
\usepackage{epsf}
\usepackage{amsmath}
\usepackage{amssymb}

\usepackage{graphicx}
\usepackage{dcolumn}
\usepackage{bm}
\def\be{\begin{equation}}
\def\ee{\end{equation}}
\def\bea{\begin{eqnarray}}
\def\eea{\end{eqnarray}}

\def\l{\lambda}

\def\m{\mu}
\def\n{\nu}

\def\s{\sigma}

\usepackage{color}

\begin{document}

\preprint{IPMU-10-0103}
\preprint{NSF-KITP-10-085}

\title{Black holes in an asymptotically safe gravity theory \\ with higher derivatives}

\medskip\
\author{Yi-Fu Cai}
\email[Email:]{caiyf@ihep.ac.cn}
\affiliation{ Institute of High Energy Physics, Chinese Academy of
Sciences, P.O. Box 918-4, Beijing 100049, China}
\affiliation{ Department of Physics \& School of Earth and Space
Exploration  \& Beyond Center, Arizona State University, Tempe, AZ
85287, USA}
\affiliation{ Institute for the Physics and Mathematics of the
Universe, University of Tokyo, 5-1-5 Kashiwanoha, Kashiwa, Chiba
277-8568, Japan}
\author{ Damien A. Easson}%
\email[Email:]{easson@asu.edu}
\affiliation{ Department of Physics  \& School of Earth and Space
Exploration  \& Beyond Center, Arizona State University, Tempe, AZ
85287, USA}
\affiliation{ Institute for the Physics and Mathematics of the
Universe, University of Tokyo, 5-1-5 Kashiwanoha, Kashiwa, Chiba
277-8568, Japan}
\affiliation{Kavli Institute for Theoretical Physics, University
of California, Santa Barbara, CA 93106-4030, USA}


\begin{abstract}
We present a class of spherically symmetric vacuum solutions to an
asymptotically safe theory of gravity containing high-derivative
terms. We find quantum corrected
Schwarzschild-(anti)-de Sitter solutions with running
gravitational coupling parameters. The evolution of the couplings
is determined by their corresponding renormalization group flow
equations. These black holes exhibit properties of a classical
Schwarzschild solution at large length scales. At the center, the
metric factor remains smooth but the curvature singularity, while
softened by the quantum corrections, persists. The solutions
have an outer event horizon and an inner Cauchy horizon which
equate when the physical mass decreases to a critical value.
Super-extremal solutions with
masses below the critical value correspond to naked
singularities. The Hawking temperature of the black hole vanishes
 when the physical mass reaches the critical value.
Hence, the black holes in the asymptotically safe gravitational
theory never completely evaporate. For appropriate values of the
parameters such stable black hole remnants make excellent
dark matter candidates.
\end{abstract}


\maketitle

\section{Introduction}

One of the most challenging tasks facing theoretical physicists
today is the construction of a consistent ultraviolet (UV)
complete theory of gravity. Weinberg has suggested that the
effective quantum field description of a gravitation theory may be
UV complete and non-perturbatively renormalizable by virtue of
asymptotic safety (AS)~\cite{Weinberg:1977}. In this scenario the
renormalization group (RG) flows have a fixed point in the UV
limit and a finite dimensional critical surface of trajectories
approach this point at short distances. This
theory has been extensively studied in the literature~\cite{Weinberg:1979,
Kawai:1993mb, Reuter:1996cp, Souma:1999at, Lauscher:2001ya,
Lauscher:2001rz, Litim:2003vp, Niedermaier:2006ns}, and recent
evidence suggests the UV critical surface is only
three-dimensional even in truncations of the exact RG equations
with more than three independent coupling
parameters~\cite{Codello:2007bd,Codello:2008vh,Benedetti:2009rx,Weinberg:2009ca,
Weinberg:2009bg, Weinberg:2009wa}.
Until now, the majority of the work on the subject has
considered significant truncations of the action, taking into account
only the Einstein-Hilbert and occasionally cosmological constant terms.
In this paper, we initiate the study of black hole solutions in an
asymptotically safe gravity theory including higher derivative terms and running
gravitational couplings. While we focus on the above goal, our methodology is easily generalized
to include a broader analysis of black solutions in higher derivative theories.

Our concrete starting point is a generally
covariant gravitational theory with
effective action involving a momentum cutoff $p$:
\begin{eqnarray}\label{action}
 \Gamma_{p}[g_{\mu\nu}]
 &=& \int{d}^4x\sqrt{-g} \bigg[ p^4g_0(p) +p^2g_1({p})R \nonumber\\
 && +g_{2a}(p)R^2 +g_{2b}(p)R_{\mu\nu}R^{\mu\nu} \nonumber\\
 &&+g_{2c}(p)R_{\mu\nu\sigma\rho}R^{\mu\nu\sigma\rho}
 + \mathcal{O} (p^{-2}R^3)+ \cdots
 \bigg]~,
\end{eqnarray}
where $g$ is the determinant of the metric tensor $g_{\m\n}$, $R$ is the Ricci scalar, $R_{\m\n}$ is the Ricci tensor and $R_{\m\n\l\s}$ is the Riemann tensor. The coefficients $g_i$ ($i = 0,\,1,\,2a,\dots$) are dimensionless coupling parameters
and are functions of the dimension-full, UV cutoff. In particular, for long wavelengths
we have
\be
g_0 (p) = -\frac{\Lambda(p)}{8 \pi G_N (p)} \, p^{-4} \,, \qquad
g_1(p) = \frac{1}{8\pi G_N(p)}\, p^{-2}\,,
\ee
where it is useful to define the dimensionless Newton's constant
$\bar{G}_N(p) = G_N(p) p^2$ and dimensionless cosmological constant
$\bar{\Lambda}(p) = \Lambda(p) p^{-2}$.
The couplings satisfy the following RG
equations,
\begin{eqnarray}\label{RGeq}
\frac{d}{d\ln{p}} \, g_i(p)=\beta_i[g(p)]~.
\end{eqnarray}

The conditions for asymptotic safety require that all the beta functions vanish
$\beta_i=0$ when the coupling parameters $g_i$ approach a fixed
point $g_{i}^*$. If $g_{i}^*=0$, the fixed point is Gaussian;
if $g_{i}^*\neq0$, the fixed point is Non-Gaussian (NG).  For the NG fixed point, all the coupling parameters are fixed,
the cutoff $p$ becomes irrelevant as $p\rightarrow \infty$,
and the theory is adequately described by a finite number of higher order
counter-terms included in the effective action. Near the fixed point we
may Taylor expand the beta functions in a matrix
form
\begin{eqnarray}\label{beta}
\beta_i[g] = \sum_j {\cal B}_{ij}(g_j-g_j^*)~~,
\end{eqnarray}
where the elements of the matrix are defined by ${\cal
B}_{ij}\equiv\frac{\partial\beta_i[g]}{\partial{g}_j}^*$ at the
fixed point. Solving the RG equations (\ref{RGeq})
in the neighborhood of the fixed point we find
\begin{eqnarray}\label{gi}
 g_i(p)=g_i^*+\sum_m e_i^n \left(\frac{p}{M_*} \right)^{v_n}~,
\end{eqnarray}
where $e^n$ and $v_n$ are the suitably normalized eigenvectors and
corresponding eigenvalues of the matrix ${\cal B}_{ij}$. Since
${\cal B}$ is a general real matrix with symmetry determined
by a particular gravity model, its eigenvalues can be either real or
in pairs of complex conjugates. As a consequence, the
dimensionality of the ultraviolet critical surface is equal to the
number of eigenvalues of the matrix ${\cal B}$, of which the real
parts take negative values. The above solution involves an arbitrary
mass scale $M_*$. By requiring the largest eigenvector of order
unity, $M_*$ is typically identified with the energy scale at which the
coupling parameters are just beginning to approach the fixed
point.

It is well known that the quantization procedure for General Relativity
leads to a non-renormalizable quantum field theory, where an infinite
number of terms have to be fixed to renormalize the standard perturbation theory. Including
higher derivative terms in the action (such as we have done in (\ref{action}))
introduces higher derivative propagators that soften the
divergences of the perturbative quantization and can result in a
perturbatively renormalizable theory \cite{Stelle:1976gc}. Unfortunately,
the higher derivative terms required, introduce new massive spin-two
degrees of freedom with wrong-sign kinetic term (i.e. negative norm
states, or  \it ghosts\rm)  \cite{Stelle:1977ry}. For
example, if we quantize this system in a canonical form the
problem with ghosts arises since the metric field and its
high-derivative term are regarded as two independent variables.
These instabilities generically render the theory
non-unitary and unstable.
However, as shown in \cite{Hawking:2001yt}, if the quantum theory
is based on a path integral which is evaluated in Euclidean space
and then Wick rotated to Lorentzian space, the path integral
can yield a theory of quantum gravity without a negative norm
state. A specific application of this approach in
inflationary cosmology provides an interesting interpretation for
metric perturbations~\cite{Clunan:2009er}.
Surprizingly, evidence is mounting that
these catastrophic ghosts may be eliminated in higher derivative theories
when the UV limit of gravity is controlled by a NG fixed point as in the
AS gravity scenario~\cite{Julve:1978xn,Salam:1978fd,Floreanini:1994yp,Benedetti:2009rx, Benedetti:2009iq}.

Because the asymptotically safe gravity theory is relevant for the
physics of high energy scales and short
distances, one is naturally lead to consider its application to early
universe cosmology.
For example, Weinberg has recently argued for the existence
of inflationary solutions complete with graceful exit in the context of the theory~\cite{Weinberg:2009wa}.
Another clear application of AS gravity is to black hole physics; in particular, to understand how the theory
modifies the conventional Schwarzschild black hole solution by taking into
account the quantum corrections naturally incorporated into the model.
Some early attempts are presented in
~\cite{Bonanno:2000ep, Bonanno:2006eu, Bonanno:2009nj,
Falls:2010he}. Their key assumption was that the leading order
quantum corrections to the black hole spacetime are captured
via a simple running of the Newtonian constant $G_N$ determined
by the renormalisation group equation for gravity;
however, the consistency of the resultant modified metric and modified
effective action was not thoroughly investigated. For simplicity, the analysis
was limited to a severe truncation of the action (\ref{action}), including only
the $g_1$ term. By truncating to the Einstein-Hilbert term (as in any $f(R)$ truncation)
the authors eliminated the four-derivative propagator for the helicity two states which
may drastically alter the AS theory and its solutions.
In this paper, we develope an effective method
of finding vacuum solutions to Einstein's equation derived from the
AS gravity with higher derivative terms. We focus on spherically symmetric spacetimes and present an
exact form of a Schwarzschild-(anti)-de-Sitter (SAdS) solution.
The thermodynamical properties of these black hole solutions are briefly discussed.

The paper is organized as follows: in \S II, we derive
the equation of motion for the AS high-derivative gravity and
study a generic vacuum solution to this theory which preserves
static spherical symmetry. We find that, to leading order, this
solution is exactly SAdS. Due to the quantum nature
of renormalization, the coupling coefficients vary with respect to
the running of energy scale and it is necessary to investigate the RG
flow of these coupling coefficients in accordance with the
asymptotically safe scenario (see, \S III).  In \S IV, we
analyze the black hole solutions to the theory. We determine the scale identification between the
momentum cutoff and the radial coordinate and subsequently, we study
the behavior of the background geometry in the UV and infrared (IR) limits.
We find that the metric factor is nonsingular for all values of the radial
coordinate but the
curvature diverges at the origin. Moreover, there exists a critical mass for the
background geometry. When the black hole mass is heavier than the critical value the
black hole has two
horizons, but this solution vanishes when the mass is subcritical, revealing a naked singularity.
Numerical calculations
are performed to confirm our analytic results.
The black hole thermodynamics are briefly discussed and we
present our concluding remarks in \S V. \bigskip
\section{Static  spherically symmetric vacuum solutions}
Black hole physics provides a window into the quantum nature of
gravity. Black holes possess many remarkable properties, for example, the associated
thermodynamics~\cite{Hawking:1974sw} and
holographic properties. The simplest black hole solution in four dimensions
corresponds to a Schwarzschild spacetime~\cite{Schwarzschild:1916ae,Schwarzschild:1916uq}. As is well known, the
geometry is divided into
two causally independent regions by an event horizon located at the Schwarzschild radius
$r_s=2G_NM$~\footnote{Here and throughout we use Natural units with $c = \hbar  = k_B = 1$.}.
Given the possibility of an asymptotically safe gravitational theory,
it is logical to re-examine these intriguing properties of black hole physics when quantum
corrections are
incorporated into the gravitational action. We now obtain the field equations from the action  (\ref{action}) and search for static, spherically symmetric
solutions to the theory.

The generalized Einstein field equations are obtained by varying
the action (\ref{action}) with respect to the metric tensor $g_{\m\n}$:
\begin{eqnarray}\label{eomHD}
\tilde{G}^{\mu\nu}\equiv\frac{\delta\Gamma_{p}[g_{\mu\nu}]}{\delta{g}_{\mu\nu}}=T^{\mu\nu}~,
\end{eqnarray}
where $\tilde{G}^{\mu\nu}$ is the generalized Einstein tensor and
$T^{\mu\nu}$ is the energy-momentum tensor of the background
matter and is vanishing due to our vacuum ansatz. The
generalized Bianchi identity is preserved automatically due to
the general covariance of the scalar-type action.
The form of
$\tilde{G}^{\mu\nu}$ appearing in the above equation can be
expressed as,
\begin{widetext}
\begin{small}
\begin{eqnarray}\label{tildeG}
\tilde{G}^{\mu\nu} &=&\frac{1}{2}g^{\mu\nu}\bigg(p^4g_0+p^2g_1R+g_{2a}R^2+g_{2b}R_{\sigma\rho}R^{\sigma\rho}+g_{2c}R_{\sigma\rho\lambda\kappa}R^{\sigma\rho\lambda\kappa}\bigg)-p^2g_1R^{\mu\nu}\nonumber\\
&&+g_{2a}\bigg(-2RR^{\mu\nu}+\nabla^{\mu}\nabla^{\nu}R+\nabla^{\nu}\nabla^{\mu}R-2g^{\mu\nu}\Box{R}\bigg)\nonumber\\
&&+g_{2b}\bigg(-2R^{\mu}_{\rho}R^{\nu\rho}+\nabla_\rho\nabla^{\mu}R^{\rho\nu}+\nabla_\rho\nabla^{\nu}R^{\rho\mu}-\Box{R}^{\mu\nu}-g^{\mu\nu}\nabla^{\rho}\nabla^{\sigma}R_{\rho\sigma}\bigg)\nonumber\\
&&+g_{2c}\bigg(-2R^{\mu\rho\sigma\lambda}R^{\nu}_{\rho\sigma\lambda}-2\nabla_{\rho}\nabla_{\sigma}R^{\mu\rho\nu\sigma}-2\nabla_{\rho}\nabla_{\sigma}R^{\nu\rho\mu\sigma}\bigg)~.
\end{eqnarray}
\end{small}
\end{widetext}
In the above, we have defined the d'Alembertian
operator $\Box\equiv\nabla_{\mu}\nabla^{\mu}$, where $\nabla_{\mu}$
is the covariant derivative.  The trace of the Einstein tensor gives the useful quantity,
\begin{eqnarray}
\tilde{G}&\equiv&g_{\mu\nu}\tilde{G}^{\mu\nu}\nonumber\\
&=&2p^4g_0+p^2g_1R-2(3g_{2a}+g_{2b}+g_{2c})\Box{R}~,
\end{eqnarray}
which vanishes in our vacuum investigation,
$T^{\mu\nu}=0$. In the above derivation we have applied the
Bianchi identity:
$\nabla_{\sigma}(R^{\sigma\rho}-\frac{1}{2}g^{\sigma\rho}R)=0$.

We assume a static, spherically symmetric metric ansatz,
\begin{eqnarray}\label{metric}
 ds^2=-A(r)dt^2+\frac{dr^2}{B(r)}+r^2d\Omega_2^2~.
\end{eqnarray}
Substituting the metric (\ref{metric}) into the action (\ref{action}) and neglecting
the time integral, up to order $\mathcal{O}(p^0)$ gives,
\begin{eqnarray}\label{Gamma}
 \Gamma_{p}[A,B]
 &=& \int \! {d}r  \, 4 \pi{r}^2\sqrt{\frac{A}{B}} \,
\Big(g_0p^4 +g_1p^2R +g_{2a}R^2\nonumber\\
 && +g_{2b}R_{\mu\nu}R^{\mu\nu}
    +g_{2c}R_{\mu\nu\rho\sigma}R^{\mu\nu\rho\sigma} \Big)~,
\end{eqnarray}
where the evaluated curvature invariants in terms of the metric (\ref{metric}) are
included in the Appendix.
To obtain the vacuum solutions we must solve:
\begin{eqnarray}\label{eom}
 \frac{\delta\Gamma_{p}}{\delta{A}}=\frac{\delta\Gamma_{p}}{\delta{B}}=0~.
\end{eqnarray}
The above calculation is greatly simplified by choosing the Schwarzschild gauge
$B(r)=N(r)A(r)$ and setting $N=1$. This relation is ensured by the Cauchy
theorem for a Riemannian geometry with unique boundary.

Varying the action with respect to the function $A(r)$ gives:
\begin{eqnarray}
 \frac{\delta\Gamma}{\delta{A}}
 &=& \frac{4\pi}{r^2} \bigg[ 4(3g_{2a}+g_{2b}+g_{2c})(-2+2A-r^2A'')\nonumber\\
 &&+(2g_{2a}+g_{2b}+2g_{2c})r^3(4A^{(3)}+rA^{(4)}) \bigg]\nonumber\\
 &=&0~.
\end{eqnarray}
Solving the above equations of motion yields,
\begin{eqnarray}\label{generalsolution}
 A(r)=1+c_1r^{n_1}+c_2r^{n_2}+\frac{c_3}{r}+c_4r^2~,
\end{eqnarray}
where the exponents $n_j$ ($j = 1, \, 2$)  are given in terms of the couplings $g_{2k}$ ($k=a, \, b, \, c$),
\begin{eqnarray}
n_{1(2)}=\frac{1}{2} \left(1\mp\sqrt{\frac{50g_{2a}+17g_{2b}+18g_{2c}}{2g_{2a}+g_{2b}+2g_{2c}}} \, \right)~.\nonumber
\end{eqnarray}
This solution involves four coefficients $c_1$, $c_2$, $c_3$ and
$c_4$ which must be determined by other constraint equations,
boundary conditions and the consistency relation with infrared
limit solution.

To determine the coefficients appearing in the solution
(\ref{generalsolution}), we impose the constraint equation
$\tilde{G}=0$, consistent with our vacuum assumption. A natural choice is
to set $c_1=c_2=0$, giving the SAdS solution to the
high-derivative gravity theory:
\begin{eqnarray}\label{solution}
A(r)=1-\frac{2G_{p}M}{r}-\frac{r^2}{l_{p}^2}~,
\end{eqnarray}
where $G_{p}$ is a gravitational coupling which is varying along
with the running of the cutoff scale. The parameter $M$ is an
integration constant we identify with the physical mass
of the black hole, and $l_{p}$ is the radius of the asymptotic
(A)dS space which is also a cutoff dependent function,
expressed as
\begin{eqnarray}\label{lp2}
l_{p}^2\simeq-\frac{6g_1}{g_0p^2}\left[1+\sqrt{1-\frac{g_0}{3g_1^2} \left(12g_{2a}+3g_{2b}+2g_{2c} \right)} \, \right]
~.
\end{eqnarray}
The solution is consistent with the
usual vacuum solution from GR when the energy scale flows to the IR limit,
but differs significantly at high energy scales, as we shall see. We require a
scale identification between the coefficients of the solution and
the cutoff scale $p$ which is dependent on the details of the AS
gravity theory. Therefore, in the subsequent section we apply the AS scenario of the high-derivative
gravity action.

\section{Asymptotically safe high-derivative gravity}

The functional gravitational RG equation is based on a momentum
cutoff for the propagating degrees of freedom and captures the
nonperturbative information about the gravitational theory. The RG equation is of the
form~\cite{Reuter:1996cp}:
\begin{eqnarray}
 \frac{\partial}{\partial{\ln{p}}}\Gamma_{p}
 = \frac{1}{2}{\rm{Tr}}\bigg({\delta^{(2)}\Gamma_{p}}+{\bf{R}}(p)\bigg)
   \frac{\partial}{\partial{\ln{p}}}{\bf{R}}(p)~,
\end{eqnarray}
where ${\bf R}$ is an appropriately defined momentum cutoff at the
scale $p$ and is usually determined by the so-called optimized
cutoff process~\cite{Litim:2000ci, Litim:2001up}. In the above
formula, we suppose the gauge fixing terms have already been
included, although they are irrelevant for our present consideration.
Additionally, the trace in the RG equation
stands for a sum over spacetime indices and a loop integration. Our
philosophy then is to effectively integrate out the high momentum fluctuations with
momentum larger than the cutoff $p$, and incorporate them via the modified
dynamics for the fluctuations having momentum less than $p$.

The high-derivative terms in the effective action (\ref{action})
can be regrouped as follows,
\begin{eqnarray}\label{HDaction}
\Gamma_{p}^{\rm HD}=\int
d^4x\sqrt{-g}\bigg[\frac{\omega}{3\lambda}R^2-\frac{1}{2\lambda}C^2+\frac{\theta}{\lambda}E\bigg]~,
\end{eqnarray}
where
\begin{eqnarray}
C^2\equiv C_{\mu\nu\rho\sigma}C^{\mu\nu\rho\sigma}= R_{\mu\nu\rho\sigma}R^{\mu\nu\rho\sigma}-2R_{\mu\nu}R^{\mu\nu}+\frac{R^2}{3}~,
\end{eqnarray}
is the square of the 4-dimensional Weyl tensor, and
\begin{eqnarray}
E=R_{\mu\nu\rho\sigma}R^{\mu\nu\rho\sigma}-4R_{\mu\nu}R^{\mu\nu}+{R^2}~,
\end{eqnarray}
is the integrand of the Gauss-Bonnet term which is topological in
4-dimensional spacetime. The re-expressed high-derivative terms in
Eq. (\ref{HDaction}) are equivalent to the terms appearing in the
original action (\ref{action}), under the following identifications,
\begin{eqnarray}\label{g2abc}
 g_{2a}&=&-\frac{1}{6\lambda}+\frac{\theta}{\lambda}+\frac{\omega}{3\lambda}~,\nonumber\\
 g_{2b}&=&\frac{1}{\lambda}-\frac{4\theta}{\lambda}~,\nonumber\\
 g_{2c}&=&-\frac{1}{2\lambda}+\frac{\theta}{\lambda}~.
\end{eqnarray}

Note that a standard derivation of dimensional regularization with
$d=4-\epsilon$~\cite{Avramidi:1985ki, deBerredoPeixoto:2004if}
indicates that the beta functions of the dimensionless
coefficients introduced in (\ref{action}) are sensitive to the
dimensional corrections for the case, $\epsilon<0$. Therefore, we restrict
our attention to the case of $\epsilon\geq0$. Explicitly in
the nontrivial limit of $\epsilon\rightarrow 0$, the beta
functions for the dimensionless coefficients of the high-derivative
gravitational terms are given by~\cite{Avramidi:1985ki,
deBerredoPeixoto:2004if, Codello:2006in},
\begin{eqnarray}
 \beta_{\lambda}&=& -\frac{1}{(4\pi)^2}\frac{133}{10}\lambda^2~,\\
 \beta_{\omega} &=& -\frac{1}{(4\pi)^2}\bigg(\frac{5}{12}\lambda+\frac{183}{10}\lambda\omega+\frac{10}{3}\omega^2\bigg)~,\\
 \beta_{\theta} &=&
 -\frac{1}{(4\pi)^2}\bigg(-\frac{196}{45}\lambda+\frac{133}{10}\lambda\theta\bigg)~.
\end{eqnarray}
To perform the stability analysis, we note that the
coefficient $\lambda$ has the familiar logarithmic form which
approaches asymptotic freedom,
\begin{eqnarray}
\lambda(p)=\frac{\lambda_0}{1+\frac{133}{(4\pi)^210}\lambda_0\ln{p/M_p}}~,
\end{eqnarray}
where $\lambda_0$ is a fixed value of the coefficient $\lambda$ at
the Planck scale. Since the above form is logarithmic, we find
$\lambda\simeq\lambda_0$ in a wide range around the Planck scale.
We will apply this approximation in deriving the coordinate
dependent cutoff scale later. Moreover, the other two parameters $\theta$ and $\omega$, approach
a group of fixed points in the UV limit, among which the stable ones
take the values
\begin{eqnarray}
\theta^{*}\simeq0.327~,~~\omega^{*}\simeq-0.0228~,
\end{eqnarray}
where the superscript ``$*$" denotes the parameter value at the NG
fixed point as introduced in the beginning of this paper.

By solving the beta functions for the gravitational
coupling and cosmological constant, we observe a Gaussian fixed
point in the IR limit and a NG fixed point in the UV limit. The
central result is
\begin{eqnarray}
 g_0&\simeq&-\frac{({\Lambda}_{IR}+\eta{p^2}G_{N})(1+\xi{p^2}G_{N})}{8\pi{p}^4G_{N}}~,\\
 g_1&\simeq&\frac{1+\xi{p^2}G_{N}}{16\pi{p}^2G_{N}}~,
\end{eqnarray}
where $G_{N}$ and ${\Lambda}_{IR}$ are the values of the
gravitational coupling and the cosmological constant in the IR
limit which are determined by astronomical observations.
In the UV limit, the coefficients of the
Einstein-Hilbert part of the gravity action will flow to a NG
fixed point with $g_0\rightarrow{g}_0^*$ and
$g_1\rightarrow{g}_1^*$.
To obtain the NG fixed point values of the remaining variables in the
high-derivative terms we solve the
flow equation explicitly
\begin{eqnarray}
 g_0^* &\simeq& -\frac{\eta\xi{G}_{N}}{8\pi} \simeq-6.331\times10^{-3}~,\\
 g_1^* &\simeq& \frac{\xi}{16\pi} \simeq1.432\times10^{-2}~,
\end{eqnarray}
by a numerical computation. Notice that the running gravitational
coupling $G_p$ is related to the coefficient $g_1$ by,
\begin{eqnarray}\label{Gp}
G_p=\frac{1}{16\pi{p}^2g_1}=\frac{G_N}{1+\xi{p^2}G_N}~.
\end{eqnarray}
From the analysis of $g_1$, we see that $G_p$ coincides with the
Newtonian constant $G_N$ at low energy scales but decreases
rapidly as the momentum cutoff $p$,
which implies a weakening of gravity at high energy scales. We will see
that this weakening of the gravitational force in the UV leads to
a softening of the singular behavior of the black hole solution
near the origin.

\section{Black holes \& asymptotically safe gravity}

From the above analysis, we find that the spherically symmetric
vacuum solutions to the high-derivative gravity theory flow to the
classical SAdS geometry at low energy scales. However, in order to
implement quantum corrections to the running coefficients
appearing in the classical geometry, we must determine the
relationship between the momentum cutoff $p$ and the radial
coordinate $r$.

\subsection{Relevant scale}

Recall the quantity $\tilde{G}$, introduced in Eq. (\ref{tildeG}),
vanishes for the vacuum state, $T_{\m\n} =0$. Inserting the
solution (\ref{solution}) into Eq. (\ref{tildeG}) and making use
of the redefinition (\ref{g2abc}), we could, in principle, solve
for the momentum cutoff $p$ as a function of the radial coordinate
$r$ by requiring $\tilde{G}=0$, which gives
\begin{eqnarray}
\tilde{G}=2p^4g_0+p^2g_1R-\frac{2\omega}{\lambda}\Box{R}=0~.
\end{eqnarray}
Note that the above equation is a fourth-order differential
equation and will yield a class of solutions for the momentum
cutoff $p(r)$. Consequently, it is necessary to impose the relevant
constraints in order to determine the physical solution. We now
focus our attention to the black hole solutions at high energy
scales and at small radial distances compared with the Planck scale,
$l_p$.

As a consequence of requiring  $\tilde{G}=0$, we derive a scale
identification between the momentum scale $p$ and the coordinate
$r$ of the form,
\begin{eqnarray}\label{pAS}
p(r)\simeq2.663\left(\frac{M^2}{|\lambda_0|}\right)^{\frac{1}{8}}r^{-\frac{3}{4}}~,
\end{eqnarray}
when the energy scale is as high as the AS scale. We note that
this relation is different from the result $p\sim{r}^{-3/2}$
appearing in~\cite{Falls:2010he} through a UV matching. The
discrepancy is due to the action truncation considered
in~\cite{Falls:2010he}, which consisted of only the
Einstein-Hilbert term and a vanishing cosmological constant. In
the present analysis, we include the higher derivative
terms and nonzero cosmological constant in addition to the EH term
and running $G_N$. Consequently, we consider the effective action
in the low energy limit. When the momentum cutoff flows to the IR
regime, the high-derivative terms are suppressed automatically and
thus becomes negligible. In this case,  $p\sim1/r$ which is
consistent with the IR matching of~\cite{Falls:2010he}. The
asymptotically safe SAdS geometry is obtained by replacing the
classical gravitational constant $G_N$ and the (A)dS radius with a
RG improved $G_p(r)$ and $l_p(r)$, respectively.

Note that the dS radius $l_p$ has significant cosmological
implications and may be related to the nearly
exponential acceleration of our universe at
early times, as well as, the cosmological acceleration occurring today. The running of $l_p$ can drive an
early inflationary period, complete with a successful graceful
exit~\cite{Weinberg:2009wa}. For late time cosmological
acceleration, we may consider the possibility that the value of
$l_p$ in the IR limit is of order of the size of our universe in
accordance with the latest cosmological observations.
Such cosmological applications are beyond the scope of the present
analysis.

\subsection{Black hole solutions}

Let us focus our interests on the local
quantities of the quantum corrected black hole solutions
(especially near the horizon and the origin) and leave the
investigation of dS radius to future study. One can finely tune a
very small value of $\lambda_0$ in order to let the dS radius be
much larger than the black hole horizon. In the following we study
this geometry along with the momentum cutoff $p$ flowing from the
UV regime to the IR regime.

\subsubsection{UV limit}

As the radial coordinate approaches the origin, the energy scale
of the cutoff could be very high, and consequently the coupling
parameters may already have arrived at their NG fixed point.
Correspondingly, the gravitational coupling is expressed as
\begin{eqnarray}\label{Gpo}
G_p(r)=\frac{1}{16\pi{p}^2(r)g_1}\simeq\frac{1}{\xi p^2(r)}~,
\end{eqnarray}
when $p\gg{M}_p$. In this case, the coefficient $\lambda$
approaches zero, leading to infinitely large $g_2$. According
to Eq. (\ref{lp2}), the radius of the asymptotical dS space
diverges. Therefore, we can neglect the last term of the solution
(\ref{solution}) near the origin. Inserting Eqs. (\ref{pAS}) and
(\ref{Gpo}) into (\ref{solution}), we derive the approximate form
of the metric factor in the UV limit:
\begin{eqnarray}
 A_{UV}(r)\simeq1-\frac{625}{512\pi}|\lambda_0|^{\frac{1}{4}}(Mr)^{\frac{1}{2}}~.
\end{eqnarray}

The above result indicates that the metric factor $A(r)$, unlike
its GR counterpart, is no longer singular inside the horizon of
the black hole in AS gravity with leading order
high-derivative terms. Naively, it may appear that the geometry is
glued to an asymptotical Minkowski spacetime near the origin.
However, substituting the solution into the expression for the
Ricci scalar, we obtain
\begin{eqnarray}
 R\simeq\frac{9375}{2048\pi}
 \frac{M^{\frac{1}{2}}|\lambda_0|^{\frac{1}{4}}}{r^{\frac{3}{2}}} ~,
\end{eqnarray}
which is singular at the origin of the spacetime. By applying
the horizon condition $A_{UV}(r)=0$, we obtain an
approximate solution
\begin{eqnarray}\label{rUV}
r_{UV}\simeq\frac{0.671\pi^2}{|\lambda_0|^{\frac{1}{2}}M}~,
\end{eqnarray}
at high energy scales, indicating that  the curvature
singularity is hidden behind the UV horizon. The other familiar invariants
are altered as well and reflect the softening of the singular behavior in the asymptotically safe gravity theory:
\bea
R_{\m\n\l\s} R^{\m\n\l\s} &\simeq& \frac{31640625}{4194304} \frac{\sqrt \l_0}{\pi^2}\frac{M}{ r^3} \,, \\
C_{\m\n\l\s} C^{\m\n\l\s} &\simeq& \frac{1171875}{4194304} \frac{\sqrt \l_0}{\pi^2}\frac{M}{ r^3}  \,, \\
R_{\m\n}R^{\m\n} &\simeq& \frac{59765625}{8388608} \frac{\sqrt
\l_0}{\pi^2}\frac{M}{ r^3} \,.
\eea
These should be compared with
their hard  singularity partners in General Relativity, for example, the Kretschmann scalar (37) in GR
diverges at the origin as $\sim 1/ r^6$.

\subsubsection{IR limit}

Along with the increasing of the radial coordinate the momentum
cutoff could drop to the regime below the Planck scale as
$p\sim1/r$. In this case,
\begin{eqnarray}
G_p(r)\simeq{G_N} \left(1-\frac{\tilde\xi{G_N}}{r^2} \right)~,
\end{eqnarray}
where $\tilde\xi$ is a constant which deviates from the
coefficient $\xi$ by a factor of $\mathcal{O}(1)$. Substituting
this gravitational coupling parameter into the black hole solution
yields
\begin{eqnarray}
A_{IR}(r)\simeq1-\frac{2G_NM}{r} \left(1-\frac{\tilde\xi{G_N}}{r^2} \right)~.
\end{eqnarray}

By solving the equation of horizon condition, we obtain another
real solution which corresponds to the event horizon at low energy
scales. Expanding to the leading order in the coefficient
$\tilde\xi$, we obtain the familiar Schwarzschild horizon value,
\begin{eqnarray}\label{rIR}
r_{IR}\simeq
2G_NM-\frac{\tilde\xi}{2M}+\mathcal{O}(\tilde\xi^2)~. 
\end{eqnarray}
Hence, the horizon location of the AS black hole is in
precise agreement with the usual form  for the Schwarzschild radius $r_s = 2G_NM$ in the limit
$\tilde\xi\rightarrow0$.

By comparing the solution for the event horizon in the UV
limit (\ref{rUV}) and the horizon in the IR limit (\ref{rIR}), we
see there is a critical value for the mass parameter
of the black hole:
\begin{eqnarray}\label{Mc}
 M_c\sim|\lambda_0|^{-\frac{1}{4}}G_N^{-\frac{1}{2}}~.
\end{eqnarray}
When $M>M_c$, the black hole in AS high-derivative gravity has two
horizons.  There is an inner Cauchy horizon as well as an outer
event horizon corresponding to that of the ordinary black hole
solution in GR. At the extremal value $M=M_c$, these two horizons
coincide; beyond the extremal value ($M<M_c$ ) the black hole
vanishes, corresponding to a naked singularity. The double horizon
feature of our solution is in agreement with the results obtained
in the AS gravity with Einstein truncation analyzed
in~\cite{Bonanno:2000ep}.

Here we make further comparison between
our results and those obtained in Refs.
\cite{Bonanno:2000ep, Bonanno:2006eu, Bonanno:2009nj,
Falls:2010he}. First, due to the presence of
higher derivative terms, the form of the scale identification (32)
is altered compared with the result obtained in the case of the pure Einstein
truncation~\cite{Falls:2010he}. Furthermore,
the black hole solutions obtained with Einstein
truncation are smooth and nonsingular, but become singular if
higher derivative terms are taken into account. The
conditions identifying the critical mass are the same in the two
scenarios, i.e., both the metric factor and its derivative with
respect to the mass are vanishing.

\subsubsection{Numerical analysis}

\begin{figure}[htbp]
\includegraphics[scale=0.8]{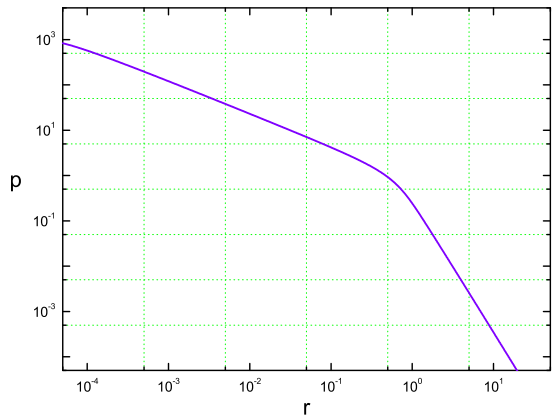}
\caption{Running of the momentum cutoff $p$ as a function of the
radial coordinate $r$. Here we
take $\lambda_0=G_N=1$.} \label{fig:cutoff}
\end{figure}

\begin{figure}[htbp]
\includegraphics[scale=0.8]{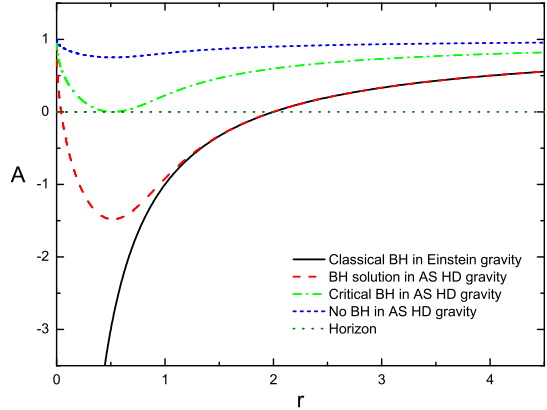}
\caption{The metric factors $A$ of the spherically symmetric
vacuum solutions as functions of the radial coordinate $r$ in both
Einstein gravity and the AS high-derivative gravity. In the
numerical computation, we take $\lambda_0=G_N=1$. We consider
$M=1$, $M=0.403$ and $M=0.1$, which correspond to a RG running
black hole, a critical black hole, and a naked singularity,
respectively.} \label{fig:metric}
\end{figure}

In the above analytic study we made a number of approximations in
order to obtain the black hole solutions in the AS high-derivative
gravity theory. To better understand the analytic results, we now
turn to a quantitative numerical analysis. Our results for the
coordinate-dependent momentum cutoff $p(r)$ and the quantum
corrected Schwarzschild metric factor $A(r)$ are presented in
Figs. \ref{fig:cutoff} and \ref{fig:metric}, respectively.

Fig. \ref{fig:cutoff},  shows the slope of the running momentum
cutoff curve in the IR and UV regimes. In the IR regime, which
corresponds to a large length scale, $p$ scales approximately as
$1/r$. However, in the high energy limit, the relation between the
momentum cutoff and the radial coordinate becomes $p\propto
r^{-3/4}$ which is in agreement with the analytic result obtained
in Eq. (\ref{pAS}).

In Fig. \ref{fig:metric}, we plot the $g_{00}$ metric component of
the spherically symmetric vacuum solutions of AS high-derivative
gravity for the values of the mass parameter: $M=1$, $M=0.403$ and
$M=0.1$. In order to illustrate the significance of the quantum
corrections to the vacuum solutions, we compare them with the
metric factor of a black hole in classical Einstein gravity as
denoted by the black solid curve. The red dash curve corresponds
to the metric factor of a double-horizon black hole with $M=1$ in
AS high-derivative gravity. When the physical mass decreases to
the critical value ($M_c=0.403$ in this example), we obtain an
extremal black hole and the outer horizon coincides with the inner
horizon. This case is plotted by the green dash-dot curve.
Moreover, the blue short dash curve shows that there are no black
hole solutions (i.e. there is a naked singularity) for subcritical
values of the mass $M<M_c$. In all of the above solutions the
metric factors of the background geometries are smooth and
nonsingular, but the curvature is divergent at the origin.

\subsection{Thermodynamics of the IR horizon}

From the above analysis, we have shown how the normal
Schwarzschild geometry is corrected by nonperturbative effects in
quantum gravity. To further probe the quantum gravity effects we
now study the process of black hole evaporation via Hawking
radiation~\cite{Hawking:1974sw}. We apply the Euclidean path
integral approach to determine the temperature and the specific
heat~\cite{Gibbons:1976ue}, and employ the method of complex path
analysis to study the emission rate of the Hawking
radiation~\cite{Srinivasan:1998ty, Parikh:1999mf}. We now
investigate the thermodynamical features of the quantum corrected
black hole around the IR horizon in the super-critical solution
with $M \geq M_c$.

\begin{figure}[htbp]
\includegraphics[scale=0.8]{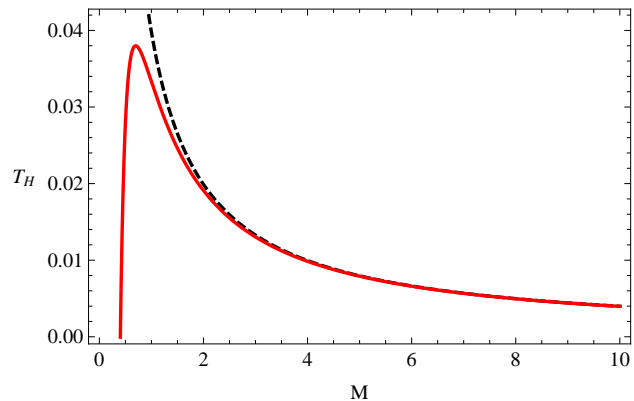}
\caption{Approximate Hawking temperature as a function of black
hole mass (see Eq.~(\ref{htemp})). The AS high-derivative solution
is the solid (red) curve and is compared with the standard Hawking
temperature of General Relativity $T^{-1} = 8 \pi M$, dashed
(black) curve, which diverges for small $M$.   In the numerical
computation, we take $\tilde \xi =.65$ and $G_N=1$. The
temperature reaches a maximum at $M = \sqrt{3 \tilde \xi/4 G_N}$
corresponding
 to $T^{-1}_{max} = 6 \pi \sqrt{3 G_N \tilde \xi} $, and reaches zero at
 the critical value $M_c = \sqrt{\tilde \xi/4G_N} \simeq .403$.}
\label{fig:htemp}
\end{figure}

There has been considerable interest in the study of
thermodynamics of quantum corrected black holes. For example, the
Hawking radiation of a two dimensional nonsingular black
hole~\cite{Easson:2002tg} in the context of dilaton
gravity~\cite{Trodden:1993dm, Mukhanov:1991zn}. Similar to the
derivation performed in~\cite{Easson:2002tg}, we compute the
Bekenstein-Hawking temperature
\begin{eqnarray}\label{htemp}
 T &=& \frac{1}{4\pi}A'(r_{IR}) \nonumber\\
   &\simeq& \frac{1}{8{\pi}G_NM} \left(1-\frac{\tilde\xi}{4 G_NM^2} \right) ~,
\end{eqnarray}
around the IR horizon. In comparison with the standard result, the
temperature of the quantum corrected black hole is smaller than that
of the classical analog (see Fig.~\ref{fig:htemp}).
Stefan's law gives the black hole radiation power law derivation:
\begin{eqnarray}
 {\cal P} &=& -\frac{dM}{dt}=4\pi\sigma{r_{I \! R}^2}T^4 \nonumber\\
          &\simeq& \frac{\sigma}{256\pi^3G_N^2M^2} \left(1-\frac{3\tilde\xi}{2G_NM}  \right)~,
\end{eqnarray}
where $\sigma=\pi^2/60$ is a Stefan-Boltzmann constant. Hence, the
evaporation of the quantum corrected black hole can only take
place when its physical mass is heavier than the critical value.
Consider $\lambda_0\sim \mathcal{O}(1)$, this critical value
coincides with the critical mass $M_c$ as introduced in the
previous subsection. Once the physical mass is smaller than $M_c$,
the absorption process will dominate over the emission process of
the black hole radiation. Therefore, a critical mass black hole
with $M\sim M_c$ is expected to be the most stable. Indeed, from
Fig.~(\ref{fig:htemp}) we see that the extremal black hole ceases
to decay because the temperature vanishes for the critical mass.

From the radiation power law obtained above, it is possible to
estimate the evaporation time of the black hole taking into
account the quantum gravity corrections
\begin{eqnarray}
 \tau\sim\frac{M}{|{\cal
 P}|}\sim\frac{\tau_{S}}{1-\frac{3\tilde\xi}{2G_NM}}~,
\end{eqnarray}
where $\tau_{S}$ is the evaporation time of a classical black
hole. The quantum corrected evaporation time is infinite when the
physical mass of the black hole decreases to the scale around the
critical mass $M_c$, in accordance with our expectations.

In the above analysis we studied the black hole thermodynamics in
the framework of the AS high-derivative gravity theory using
perturbative methods with $\tilde\xi<1$. However, we do not expect
these qualitative results to be drastically altered in the
nonperturbative regime. This is due to the fact that there is
always a critical mass for the black hole solution where its IR
horizon coincides with the UV horizon. This corresponds to an
extremal black hole with a vanishing temperature (since $T\propto
A'=0$). As a consequence, the black hole evaporation time will be
infinite as discussed above.

\section{Conclusions}\label{conc}

In this paper we have initiated the study of static spherically
symmetric vacuum solutions to the theory of asymptotically safe
gravity with high-derivative corrections. We find that a generic
solution corresponds to a quantum corrected SAdS spacetime due to
the RG flows of the gravitational coupling parameters. Under
certain simplifying assumptions, we obtain a black hole solution
with smooth metric factor and a curvature singularity at the
center of the geometry. The singular behavior is mildly softened
in the AS high derivative theory compared to the corresponding GR
black hole. The quantum gravity corrected black hole solution
generically possesses two horizons, which correspond when the
physical mass of the black hole decreases to a critical value. The
temperature of the black hole is exactly zero when its physical
mass reaches the critical value. We find that the temperature of
the quantum corrected black hole is, in general, lower than that
of a classical black hole and stable critical mass remnants  are
natural final states of the Hawking evaporation process.

These results are related to many other interesting issues which
deserve future study. For example, a collection of critical black
holes may serve as an alternative candidate for dark matter if
their masses are as low as the ${\rm TeV}$ scale. From Eq.
(\ref{Mc}), we find that this is a possibility if
$\lambda_0\sim10^{64}$ which is also compatible with the
theoretical requirement of suppressing gravitational
high-derivative terms. The passive thermodynamic properties of the
AS black hole with a critical mass may leave significant signals
on the matter power spectrum at small length scales. In addition,
if sufficient numbers of critical mass holes were produced at the
end of inflation, they may leave potentially measurable signals in
the polarization of the cosmic microwave background radiation on
small angular scales. A potential concern is that if the critical mass of a
black hole is
much lower than the Planck scale Minkowski spacetime could become unstable due to the
perturbation modes which increase the mass of the black hole. It is then possible that
there would be too many primordial black hole remnants
produced in a static background. We note that this problem may be alleviated  in realistic
cosmological models due to the expansion of the
universe. As the expansion of the background spacetime dominates
over the production rate of black holes, the density of
primordial black holes may be significantly diluted after a sufficiently
long period. We leave the study of this intriguing possibility
for future work.

\acknowledgments
It is a pleasure to thank R.~Brandenberger, R.~Emparan,
K.~Stelle and S.~Weinberg for helpful conversations. YFC
acknowledges Prof. Xinmin Zhang for extensive support of his
research. YFC thanks the Institute for the Physics and Mathematics
of the Universe and the Research Center for the Early Universe at
the University of Tokyo, Tokyo University of Science, and the
Yukawa Institute for Theoretical Physics at Kyoto University for
their hospitality when this work was finalized. The work of YFC is
supported in part by the National Science Foundation of China
under Grants No. 10533010 and 10803001. DAE
would like to thank the Yukawa Institute for Theoretical Physics at
Kyoto University for
their hospitality during YKIS2010 (YITP-T-10-01) where this
work was completed.
The work of DAE is
supported in part by the World Premier International Research
Center Initiative (WPI Initiative), MEXT, Japan and by a
Grant-in-Aid for Scientific Research (21740167) from the Japan
Society for Promotion of Science (JSPS), and by funds from the
Arizona State University Foundation, and by the National Science
Foundation under Grant~No.~PHY05-51164.

\newpage
\appendix*
\section{Curvature invariants}

\begin{widetext}

The curvature invariants appearing in the action (\ref{action})
evaluated in terms of the line element (\ref{metric}):
\be R =
\frac{4 A^2-4 A^2 B-4 r A B A'+r^2 BA'^2-4 r A^2 B'-r^2 AA'B'-2
r^2 AB A''}{2 r^2 A^2}
\ee

\bea
R^2 &=& \frac{1} {4 r^4 A^4} \big(-(r^2 BA'^2) +  r A \left( 2 r B A'' + A' \left( r B'+4 B \right) \right) \nonumber \\
&+& 4 A^2 \left( r B'+B-1 \right) \big)^2
\eea

\bea
R_{\m\n} R^{\m\n} &=& \frac{1}{8 r^4 A^4}(r^4 B^2 A'^4+4 A^4(4 (-1+B)^2+4 r (-1+B) B'+3 r^2 B'^2) \nonumber \\
&+& 4 r A^3(A'(4 (-1+B) B+2 r BB'+r^2 B'^2)+2 r^2 B B' A'') \nonumber \\
&+& r^2 A^2(A'^2(12 B^2+r^2 B'^2) + 4 r B A'(2 B+r B') A''+4 r^2 B^2 A''^2) \nonumber \\
&-& 2 r^3 A BA'^2(r A' B'+2 B(A'+r A'')))
\eea

\bea
R_{\m\n\l\s} R^{\m\n\l\s} &=& \frac{1}{4 r^4 A^4}(r^4 B^2 A'^4+8 A^4(2 (-1+B)^2+r^2 B'^2)\nonumber \\
&+& -2 r^4 A B A'^2(A'B'+2 B A'') +  A^2(r^2 A'^2(8 B^2+r^2 B'^2) \nonumber \\
&+& 4 r^4 B A'B'A''+4 r^4 B^2 A''^2))
\eea

\end{widetext}

\end{document}